\documentclass{aa}

\usepackage[varg]{txfonts}
\usepackage[english]{babel}
\usepackage[utf8]{inputenc}
\usepackage{graphicx}
\usepackage{amsmath}
\usepackage{units}
\usepackage{tikz}

\bibpunct{(}{)}{;}{a}{}{,} 

\title{On the origin of intrinsic alignment in cosmic shear
  measurements: an analytic argument}
\author{Giovanni~Camelio\inst{\ref{inst1},\ref{inst2}} \and
Marco~Lombardi\inst{\ref{inst1},\ref{inst3}}}
\authorrunning{G. Camelio \and M. Lombardi}
\offprints{giovanni.camelio@roma1.infn.it}
 \institute{ University of Milan, Department of Physics, via Celoria 16,
 I-20133 Milan, Italy\label{inst1} \and University of Rome ``La Sapienza'', Department of
 Physics, p.le Aldo Moro 2, I-00185 Rome,
 Italy\mail{giovanni.camelio@roma1.infn.it}\label{inst2} \and Harvard-Smithsonian Center for
 Astrophysics, 60 Garden Street, Cambridge, MA 02138, USA\label{inst3}}
\date{Received ***date***; Accepted ***date***}
\keywords{gravitational lensing: weak -- galaxies: kinematics and dynamics --
galaxies: halos}

\begin{document}

\abstract{Galaxy intrinsic alignment can be a severe source of error in weak-lensing studies. The problem has been widely studied by numerical simulations
and with heuristic models, but without a clear theoretical justification of its
origin and amplitude. In particular, it is still unclear whether intrinsic
alignment of galaxies is dominated by formation and accretion processes or by
the effects of the instantaneous tidal field acting upon them. We investigate
this question by developing a simple model of intrinsic alignment for elliptical
galaxies, based on the instantaneous tidal field. Making use of the galaxy stellar
distribution function, we estimate the intrinsic alignment signal and find that
although it has the expected dependence on the tidal field, it is too
weak to account for the observed signal. This is an indirect validation of the
standard view that intrinsic alignment is caused by formation and/or
accretion processes.}

\maketitle

\section{Introduction}

It has long been recognized that gravitational lensing is a particularly robust
method to investigate the mass distribution of massive astronomical objects
\citep[e.g., see][]{article--Refsdal--1964--gravitational,
article--Fu_al--2008--very, article--Jauzac_al--2012--weak-lensing}.
Gravitational lensing is insensitive to the chemical and physical state of the
deflecting matter and therefore treats ordinary
and dark matter components of the deflector equally.  In addition, it depends on a clearly
understood physical process, the distortion of the space-time induced by masses
as described by General Relativity, for which it is not required that the
deflector be in a state of dynamical equilibrium.  As a result, it
currently represents our best tool to study the matter distribution in the Universe.

A particularly important use of weak-lensing methods, which are
based on the statistical analysis of the weak correlation among the observed
shapes of distant galaxies, is studying the large-scale structure of the Universe.
In \textup{weak cosmological lensing}, the deflecting matter
studied (in contrast to the case of a cluster of galaxies) is
distributed along the entire line of sight.  As a result, the distinction
between source and lens is less clear: the light of every galaxy that
contributes to the distortion of space-time is bent by the distribution of
matter along its path.  The correlation between the apparent ellipticities of
every pair of galaxies determines the signal; this is called
\textit{\textup{cosmic shear}}.  Since cosmic shear is related to
the power spectrum of the density contrast \citep[e.g., see][]
{article--Miralda-Escude--1991--correlation, article--Kaiser--1992--weak}, with this tool one measures the statistical properties of
the large-scale structure in the Universe.
Therefore, weak cosmological lensing is an independent method for testing
cosmological models \citep[e.g., see][]{article--Kilbinger_all--2012--cfhtlens}.

Cosmic-shear measurements are particularly challenging: the signal is very weak
and is affected by many error sources.  From the modeling point of view, one of
the main difficulties is related to the so-called \emph{\textup{intrinsic alignment}}, that is,
the galaxy intrinsic shape alignment, which is not caused by gravitational lensing, but by
the source gravitational field (\citealp
{article--Heavens_Refregier_Heymans--2000--intrinsic,article--Croft_Metzler--2000--weak};
see also \citealp{Troxel+Ishak_arXiv_2014} for a recent review).  Although the
effects of intrinsic alignment can be reduced by removing pairs of galaxies that
are \textit{\textup{physically close}} to the measurement of the shear correlation
\citep{article--King_Schneider--2002--suppressing,article--Heymans_Heavens--2003--weak},
there remains a residual bias (called GI mode, see below). Therefore, it is
important to clarify these effects and to assess their impact on current and
future cosmic-shear surveys. Although some heuristic models for
intrinsic alignment are available
\citep[e.g.,][]{article--Catelan_Kamionkowski_Blandford--2001--intrinsic,
article--Hirata_Seljak--2004--intrinsic}, they are unfortunately
incomplete; in particular,
they lack an analytic estimate of the bias magnitude (in general, such an estimate
would require the use of complicated galaxy formation and stellar dynamical
models). Therefore, one has to resort to the bias magnitude measured from
simulations \citep{article--Croft_Metzler--2000--weak,
article--Heavens_Refregier_Heymans--2000--intrinsic,Jing_MNRAS335_2002,
Heymans+_MNRAS371_2006, Tenneti+_MNRAS441_2014} for a comparison with the
observations.

The exact origin of intrinsic alignment is still debated (e.g., see
\citealp{letter--Pereira_Kuhn--2005--radial}), and it is still unclear
whether intrinsic alignment of galaxies is dominated by formation and
accretion processes or by the effects of the instantaneous tidal field
acting upon them.  The commonly accepted view, supported by
simulations and observations, is that the intrinsic alignment is
caused by the tidal field acting at the formation epoch
\citep{article--Catelan_Kamionkowski_Blandford--2001--intrinsic,
  article--Hirata_Seljak--2004--intrinsic} and possibly on the merger
history.  In this paper we strengthen this view with analytic arguments using a \emph{\textup{reductio ad
  absurdum}}: that is, we start with the assumption that intrinsic
alignment is dominated by the instantaneous tidal field currently at
work on galaxies and show that the predicted effect is several orders
of magnitude weaker than the observed effect.  In our model, we consider a
spherical early-type galaxy and study the deformation of its shape to
first order after introducing a weak external tidal field. We
describe the effects of the external tidal field on the intrinsic
(i.e., unlensed) shape of the elliptical galaxy, determining the
luminous quadrupole of the galaxy by means of its collisionless stellar
distribution function. The galaxy ellipticity is then computed through
the luminous quadrupole. We find the expected dependence of the
ellipticity to the tidal field, as postulated by the heuristic model
of \citet{article--Catelan_Kamionkowski_Blandford--2001--intrinsic},
but the constant of proportionality determined by our model is more
than four orders of magnitude lower than that observed
\citep{paper--Joachimi_al--2011--constraints}, and consequently cannot account
for the intrinsic alignment.

The paper is organized as follows: in Sect.~\ref{sec:scientific-context} we
present the problem of intrinsic alignment, summarize the relevant
classifications, and outline the properties of different types of alignment.  In
Sect.~\ref{sec:dynamical-app} we introduce the stellar distribution function and
estimate the ellipticity induced by an external tidal field on an elliptical
galaxy. In Sect.~\ref{sec:discussion} we compare our result with that of
\cite{paper--Joachimi_al--2011--constraints} and apply our model to the galaxy dark
matter halo.  We draw our conclusions in Sect.~\ref{sec:conclusions}.  In
Appendix~\ref{app:rigid-body} we study the dynamics of elliptical and spiral
galaxies in the rigid-body approximation.  Finally, in Appendix
\ref{app:isopotential-app}, we derive an alternative intrinsic
alignment ellipticity determined by instantaneous tidal fields by means of the
equipotential surfaces of elliptical galaxies.  Throughout this paper we take a
standard cosmological model with Hubble constant
$H_0=\unit[70]{km\,s^{-1}\,kpc^{-1}}$, mass density parameter $\Omega_{\mathrm
m}=0.3$.

\section{General context}
\label{sec:scientific-context}

\subsection{Cosmic shear}

In principle, unbiased shear measurements can be obtained from the
luminous quadrupole tensor $\tens Q_{ij}$ of each galaxy in a given field of view
\citep[e.g.,][]{article--Bartelmann_Schneider--2001--weak}
\begin{equation}
  \label{eq:luminous-quadrupole}
  \tens Q_{ij}=\frac{\int I(\boldsymbol \theta)\theta_i\theta_j\,\mathrm
  d^2\theta}{\int
    I(\boldsymbol \theta)\,\mathrm d^2\theta} \; ,
\end{equation}
where $I(\boldsymbol\theta)$ is the surface brightness observed at
angular position $\boldsymbol \theta$, defined as the two-dimensional vector
from the luminous center of the galaxy to a given point.  In reality,
this definition of quadrupole moments produces measurements with
vanishing signal-to-noise ratio because the noise at large radii
$\theta=\|\boldsymbol \theta\|$
dominates Eq.~\eqref{eq:luminous-quadrupole}.  For this reason, in
actual weak-lensing observations one resorts to \textit{\textup{weighted}}
quadrupole tensors or to alternative measurement techniques
\citep[e.g.,][]{article--Kaiser_Squires_Broadhurst--1995--method}.

For simplicity, we use definition
\eqref{eq:luminous-quadrupole} here, which has convenient transformation
properties.  In this case
\citep{article--Schramm_Kayser--1995--complex,
article--Seitz_Schneider--1997--steps} and under the assumption that
the unlensed source ellipticities have vanishing average, the observed complex ellipticity, defined as
\begin{equation}
\label{eq:complex-ellipticity}
  \varepsilon=\frac{\tens Q_{11}-\tens Q_{22}+2\imath\tens
    Q_{12}}{\mathop{\mathrm{Tr}}\tens Q+2\sqrt{\det\tens Q}} \; ,
\end{equation}
provides an unbiased estimate of the (reduced) shear.  For a luminosity
distribution with constant flattening elliptical isophotes, the modulus of the
complex ellipticity is
\begin{equation}
  \label{eq:def-ellipticity}
  |\varepsilon| = \frac{a_\mathrm M-a_\mathrm m}{a_\mathrm M+a_\mathrm m}\;,
\end{equation}
where $a_\mathrm{M}$ and $a_\mathrm{m}$ are the semi-major and
semi-minor axes of the isophotes. If the light paths are
bent weakly, the observed ellipticity $\varepsilon$ of the galaxy is
\citep[e.g.,][]{article--Bartelmann_Schneider--2001--weak}
\begin{equation}
  \label{eq:eps-epss-gamma}
  \varepsilon\simeq\varepsilon^\mathrm s+\gamma\; ,
\end{equation}
where $\varepsilon^s$ is the intrinsic (i.e., unlensed) shape of the
galaxy and $\gamma$ is the cosmic shear.

In cosmic-shear studies it is convenient to define three
ellipticity correlation functions:
\begin{align}
  \label{eq:C1}
  C_{++}(\theta)&{}=\langle\varepsilon_+(\boldsymbol\theta_0)\varepsilon_+(\boldsymbol\theta_0
  +\boldsymbol\theta)\rangle_{\theta_0} \; ,\\
  \label{eq:C2}
  C_{\times\times}(\theta)&{}=\langle\varepsilon_\times(\boldsymbol\theta_0)
  \varepsilon_\times(\boldsymbol\theta_0
  +\boldsymbol\theta)\rangle_{\theta_0} \; ,\\
  \label{eq:C12}
  C_{+\times}(\theta)&{}=\langle\varepsilon_+(\boldsymbol\theta_0) \varepsilon_\times(\boldsymbol\theta_0
  +\boldsymbol\theta)\rangle_{\theta_0} \; ,
\end{align}
where $\varepsilon_+$ and $\varepsilon_\times$ are the real and imaginary part
of the complex ellipticity in a coordinate system with the real axis along the line connecting the two correlating
points (the centers of the two galaxies, $\boldsymbol\theta_0$ and
$\boldsymbol\theta_0+\boldsymbol\theta$). The correlation functions only depend
on the modulus of $\boldsymbol \theta$ because of isotropy.

We now consider a pair of galaxies and use the superscripts fg
and bg to denote foreground and background objects.  Then we find
\begin{multline}
\label{eq:correlazione-completa}
C_{ij}=\langle\varepsilon_i^\mathrm{fg}\varepsilon_j^\mathrm{bg}\rangle = 
\langle\gamma_i^\mathrm{fg} \gamma_j^\mathrm{bg} +
\gamma_i^\mathrm{fg}\varepsilon^\mathrm{s,bg}_j+\varepsilon_i^\mathrm{s,fg}\gamma_j^\mathrm{bg}+\varepsilon_i^\mathrm{s,fg}\varepsilon_j^\mathrm{s,bg}\rangle
\\ 
= C_{ij,\mathrm{GG}}+C_{ij,\mathrm{GI}}+C_{ij,\mathrm{II}} \; ,
\end{multline}
where $[i;j]=\{+;\times\}$, $\gamma^\mathrm{fg}$ ($\gamma^\mathrm{bg}$) is the
cosmic shear acting on the foreground (background) object, $C_{ij,\mathrm{GG}} =
\langle\gamma_i^\mathrm{fg}\gamma_j^\mathrm{bg}\rangle$ is the cosmic shear
correlation term, and $C_{ij,\mathrm{GI}}=\langle\gamma_i^\mathrm{fg}
\varepsilon_j^\mathrm{s,bg} + \varepsilon_i^\mathrm{s,fg} \gamma_j^\mathrm{bg}\rangle$
and
$C_{ij,\mathrm{II}}=\langle\varepsilon_i^\mathrm{s,fg}\varepsilon_j^\mathrm{s,bg}\rangle$
are the gravitational-intrinsic and the intrinsic-intrinsic correlation terms
(see below).  We note that the dominant term in
$C_{ij,\mathrm{GI}}$ is $\langle\varepsilon_i^\mathrm{s,fg}
\gamma_j^\mathrm{bg}\rangle$; the other term, $\langle\gamma_i^\mathrm{fg}
\varepsilon_j^\mathrm{s,bg}\rangle$, is only expected to be non-negligible if there
are long-range correlations in the gravitational tidal field (e.g., a dark
matter filament along the line of sight), and can be ignored in most
situations.  In the case of a flat cosmology, the correlation functions are
related to the density contrast power spectrum $P_\delta$
\citep[e.g., see][]{article--Bartelmann_Schneider--2001--weak,
article--Catelan_Kamionkowski_Blandford--2001--intrinsic}
\begin{align}
  C_{++,\mathrm{GG}}(\theta)+C_{\times\times,\mathrm{GG}}(\theta)&{}=\int_0^\infty\frac{\ell\,\mathrm
  d\ell}{2\pi}P_\gamma(\ell)J_0(\ell\theta)\; ,\\
  C_{++,\mathrm{GG}}(\theta)-C_{\times\times,\mathrm{GG}}(\theta)&{}=\int_0^\infty\frac{\ell\,\mathrm
  d\ell}{2\pi}P_\gamma(\ell)J_4(\ell\theta)\; ,\\
  C_{+\times,\mathrm{GG}}(\theta)&{}=0 \; ,\\
  P_\gamma(\ell)&{}=\frac{9H_0^4\Omega^2_{\mathrm m}}{4c^4}\int_0^{w_\mathrm H}\frac{W^2(w)}{a^2(w)}
  P_\delta\left(\frac \ell w,w\right)\,\mathrm dw \; ,\\
  W(w)&{}=\int_w^{w_\mathrm H} p(w')\frac{w'-w}{w'}\,\mathrm dw' \; .
\end{align}
Here $P_\gamma$ is the cosmic-shear power spectrum, $J_0$ and $J_4$
are Bessel functions of the first kind, $p(w)$ is the probability of detecting a galaxy at
comoving distance $w$, and $w_\mathrm{H}$ is the comoving horizon distance.

To obtain relevant cosmological results from cosmic-shear
measurements, it is important to understand each error source. From
the theoretical point of view, an important error source is the
contribution to the ellipticity correlation induced by the intrinsic
alignment, $C_{ij,\mathrm{GI}}+C_{ij,\mathrm{II}}$.

\subsection{Intrinsic alignment}

As noted previously in computational
\citep{article--Heavens_Refregier_Heymans--2000--intrinsic,
article--Croft_Metzler--2000--weak} and theoretical studies
\citep{article--Catelan_Kamionkowski_Blandford--2001--intrinsic},  unlensed
ellipticities need not be distributed isotropically. In particular, close pairs
of galaxies may exhibit correlated ellipticities as a result of their mutual
gravitational interaction and of an external gravitational potential.

The intrinsic alignment (IA) is the contribution to the observed galaxy
ellipticity correlation, caused by the gravitational tidal field in which
galaxies are placed. In the literature it is usually assumed
\citep{article--Catelan_Kamionkowski_Blandford--2001--intrinsic,article--Hirata_Seljak--2004--intrinsic}
that the galaxy intrinsic ellipticity is frozen at the formation epoch.  IA is
classified according to the physical generating mechanism
\citep{article--Catelan_Kamionkowski_Blandford--2001--intrinsic}, which is
different according to the type of galaxy that the
external gravitational field acts upon.

In elliptical galaxies, the external gravitational field stretches the galaxy shapes
\citep{article--Croft_Metzler--2000--weak}, partly determining their intrinsic
ellipticities.\footnote{Elliptical galaxies may be non-spherical on their own,
independently of any tidal stretching because of anisotropic pressure
and angular momentum.}  In contrast, in spiral galaxies the intrinsic ellipticity is
related to their angular momentum, produced by the torque provided by the
external gravitational field during galaxy formation and by projection effects
\citep{article--Heavens_Refregier_Heymans--2000--intrinsic}.

From a geometrical point of view \citep{article--Hirata_Seljak--2004--intrinsic}, the intrinsic correlation (for a
second-order statistics, like the shear two-point correlation function) is made
of two contributions: \emph{\textup{intrinsic-intrinsic}} (II) and
\emph{\textup{gravitational-intrinsic}} (GI).  The II contribution is the correlation of
the unlensed ellipticities of two physically close\footnote{In cosmic-shear
measurements, the II signal is usually suppressed by excluding
pairs of galaxies from the analysis that are thought to be physically close (with their physical
distance estimated from photometric redshifts and angular distance in the sky).
However, the assumption that only physically close galaxies have II alignment
might not hold because the II alignment could arise also in far away galaxies if
there are long-range correlations in the gravitational field (e.g., those
associated with a dark matter filament, the typical length scale of which is
numerically estimated as $\unit[100h^{-1}]{Mpc}$ by
\citealp{article--Springel_all--2005--simulations}; \citealp[see
also][]{article--Mandelbaum_all--2006--detection, article--Lee--2004--intrinsic,
article--Hirata_al--2007--intrinsic}).  Moreover, since the distance between
galaxies is estimated from photometric redshifts, galaxies thought to be
physically distant may actually be physically close as a result of errors in the
photometric redshift estimates.} source galaxies, generated by a correlation in
the gravitational field felt by the galaxies (which is caused by their mutual
interaction and by the external large-scale structure). In contrast, the GI
correlation is the correlation between the observed ellipticities of two
galaxies that are distant from each other, but angularly close in the sky view
plane \citep{article--Hirata_Seljak--2004--intrinsic}. This correlation is
caused by foreground mass inhomogeneities that produce two effects through the tidal
field: they induce an intrinsic ellipticity in foreground galaxies through
direct gravitational interaction and modify the \textit{\textup{observed}} ellipticities
of background galaxies by means of gravitational lensing.

\subsubsection{Elliptical galaxies}

Elliptical galaxies are expected to be polarized by the external tidal field
they are located in
\citep{article--Croft_Metzler--2000--weak,article--Catelan_Kamionkowski_Blandford--2001--intrinsic},
\begin{align}
  \label{eq:catelan1}
  \varepsilon_+^\mathrm s&{}=C(\partial_1^2-\partial_2^2)U_\mathrm{ext} \; ,\\
  \label{eq:catelan2}
  \varepsilon_\times^\mathrm s&{}=2C\partial_1\partial_2U_\mathrm{ext} \; .
\end{align}
Here $\varepsilon^s$ is the galaxy ellipticity in the plane perpendicular to the
line of sight, $U_\mathrm{ext}$ represents the large-scale potential (at the galaxy
formation epoch) associated with
inhomogeneities of cosmological origin, and $C$ is a constant
that in principle
could be determined by ``a complete galactosynthesis model''
\citep{article--Catelan_Kamionkowski_Blandford--2001--intrinsic}.
\cite{article--Catelan_Kamionkowski_Blandford--2001--intrinsic} indirectly
estimated this constant in a heuristic way by means of the relation 
\begin{equation}
  \langle\varepsilon^2\rangle=\frac8{15}C^2\left(\frac32\Omega_{\mathrm m}H_0^2\right)^2\frac1{2\pi^2}
  \int_0^\infty k^2 P_\delta(k)\left(\frac{3J_1(kR)}{kR}\right)^2\mathrm dk\; ,
\end{equation}
where $R=\unit[1h^{-1}]{Mpc}$ is the characteristic scale over which a galaxy
forms.  An II alignment for elliptical galaxies has indeed been detected in
numerical simulations \citep{article--Croft_Metzler--2000--weak} and
observations \citep[e.g.][]{letter--Pereira_Kuhn--2005--radial,
article--Agustsson_Brainerd--2006--orientation,
article--Hirata_al--2007--intrinsic, article--Faltenbacher_al--2009--alignment,
paper--Joachimi_al--2011--constraints}.

For early-type galaxy samples with broad redshift distribution and for typical
foreground mass inhomogeneities, the GI contribution is non-vanishing and is
expected to be greater than that of the II contribution
\citep{article--Hirata_Seljak--2004--intrinsic}.  The GI term is due to
the tidal effects of a mass overdensity on foreground galaxies, and the lensing
effects of the same overdensity on background galaxies. Therefore, the relevant
factor is the probability of observing an elliptical galaxy \emph{\textup{integrated}}
along the line of sight from the mass overdensity to the horizon.  In contrast,
the relevant factor for the II term is the probability of observing a second
galaxy \emph{\textup{at the same}} redshift as the first galaxy.

Finally, we note that according to this model, the GI alignment produces an anticorrelation of
ellipticities ($C_\mathrm{GI}<0$): the foreground elliptical galaxy physical (intrinsic) shape is
stretched \textit{\textup{along}} the gravitational tidal field, whereas the background galaxy apparent
shape is stretched \textit{\textup{perpendicular}} to it because of lensing
\citep{article--Hirata_Seljak--2004--intrinsic}. For
example, in a galaxy cluster, cluster members would be preferentially aligned
radially, whereas the background galaxies would be seen preferentially
tangentially.

\subsubsection{Spiral galaxies}

In spiral galaxies the II contribution is thought to be caused by the torque
provided by the external tidal field during formation
\citep{article--Heavens_Refregier_Heymans--2000--intrinsic,
article--Catelan_Kamionkowski_Blandford--2001--intrinsic,
article--Hirata_Seljak--2004--intrinsic, article--Hirata_Seljak--2010--erratum}.
This contribution is to second order in the external tidal field because the
tidal field first has to generate an anisotropic moment of inertia $\tens
I_{ij}$ (as in elliptical galaxies, the distribution of mass is expected to be
elongated along the gravitational field gradient), and then to torque it
\citep{article--Peebles--1969--origin, article--Doroshkevich--1970--space,
article--white--1984--angular}.  The torque provided by the external tidal field
generates the angular momentum of the spiral proto-galaxy (this is similar to
what we show in Appendix~\ref{app:rigid-body}, where we compute the torque
generated by the external tidal field on an elliptical galaxy not aligned with
it).
One then needs to specify the correlation between the angular momentum and the
galaxy ellipticity,
for example by assuming
zero thickness for the (circular) disk of the galaxy and considering the projection
effects, as reported by \citet{article--Heavens_Refregier_Heymans--2000--intrinsic}.
Finally, since a correlation in the tidal field induces a correlation in the angular momenta of
close spiral galaxy pairs, the expected II
power spectrum can be computed analytically, as shown by \citet{article--Hirata_Seljak--2004--intrinsic}.

The II contribution for spiral galaxies is of higher order in the tidal field
and thus should be lower than for ellipticals
\citep{article--Hirata_Seljak--2004--intrinsic}. In addition, little GI contribution
is expected for spiral galaxies \citep{article--Hirata_Seljak--2004--intrinsic}.
At present, there is no clear observational evidence of IA in spiral galaxies
\citep{article--Hirata_al--2007--intrinsic,
article--Faltenbacher_al--2009--alignment,
article--Blazek_all--2012--separating}.

In Appendix~\ref{app:rigid-body} we determine the rigid-body precession period
of spiral galaxies, which is longer than the galaxy deformation time. This shows that no
precession-driven mechanism for an alignment of spiral galaxies with the external
tidal field can be devised.

\section{Distribution function method}
\label{sec:dynamical-app}

Even if galaxies are many-body systems, it might na\"ively be
thought that their
alignment is driven by rigid-body motions (e.g., oscillations and precessions)
under the effects of external gravitational forces.  However, an analysis of the
relevant time scales shows that rigid-body motions are much slower than those
characterizing internal dynamics (Appendix~\ref{app:rigid-body},  see also
\citealp{article--Ciotti_Dutta--1994--alignment}).  This means
that at least for IA
purposes, the galaxy is best studied as a deforming body.  In this section, we
start from the hypothesis that galaxies are not subject to any intrinsic alignment during formation
and that intrinsic alignment is entirely due to tidal fields at the observation
time, to which galaxies react immediately. As shown in
Sect.~\ref{sec:discussion}, this assumption leads to an extremely low IA,
completely inconsistent with observations. This is an indirect proof that IA is
driven by formation and merging.

The deformation of an elliptical galaxy subject to an external
gravitational field can be modeled in a simplified way by studying the
deformation of its equipotential surfaces (see
Appendix~\ref{app:isopotential-app}), and in a more complete manner, by means of
the stellar distribution function. In both cases, we start with
unperturbed spherical galaxies for simplicity and study the ellipticity induced by the external
tidal field. The spherical assumption is quite strong, and it may not be straightforward
to generalize our results to the case of an \emph{\textup{ensemble average}} of
elliptical early-type galaxies. Nevertheless, since we are interested in an order-of-magnitude estimate, the spherical assumption is sufficient for our purposes.

\subsection{General case}

A galaxy is a complex many-body system, with a gravity-driven dynamics
\citep{book--Chandrasekhar--1942--principles,book--Bertin--2000--dynamics}.  In
this context, an important tool is the stellar distribution function $f_\star(\vec
x,\vec{\dot x}, t)$, which is, essentially, the stellar density distribution in
phase space.  We also define \citep[e.g.,][]{book--Bertin--2000--dynamics}
the (total) distribution function $f=f_\star+f_\mathrm{DM}$ as the sum of the
stellar distribution function and the dark matter distribution function. We assume
the collisionless Boltzmann equation and the Poisson equation:
\begin{equation}
  \label{eq:hamilton-f}
  \frac{\partial f}{\partial t}+\frac{\partial f}{\partial x_i}\dot
  x_i-\frac{\partial f}{\partial \dot x_i}\frac{\partial U}{\partial
    x_i}=0\;,
\end{equation}
\begin{multline}
  \label{eq:poisson-f}
  \nabla^2U(\vec x,t)=4\pi G\rho(\vec x,t)=4\pi G m\int f(\vec x, \vec{\dot
  x},t)\,\mathrm d^3\!\dot x\\
  =4\pi G m\int f_\star(\vec x, \vec{\dot x},t)\,\mathrm d^3\!\dot x + 4\pi
  G m\int f_\mathrm{DM}(\vec x, \vec{\dot x},t)\,\mathrm d^3\!\dot x\;,
\end{multline}
where $m$ is the mean mass of a galaxy star (and of a DM ``particle'' as well).  In the following,
we use the King model \citep{article--King--1966--structure} for the stellar
distribution function\footnote{The King model
is usually applied to study globular clusters; more complex distribution
functions are generally used for galaxies. Here, we choose the King model for
its simplicity and ability to clarify the key aspects of the
dynamical model we are interested in.}
\begin{equation}
  f_\star(\vec x,\vec{\dot x},t)=\begin{cases}A\Big[\mathrm e^{-U(\vec
      x)/\sigma_v^2-\|\vec{\dot x}\|^2/(2\sigma_v^2)}-\mathrm
                e^{-E_\mathrm{tr}/\sigma_v^2}\Big]&E<E_\mathrm{tr}\;,\\
    0&E\ge E_\mathrm{tr}\;,\end{cases}
\end{equation}
where $E=U(\vec x)+\frac{\|\vec{\dot x}\|^2}{2}$ is the single-star energy and
$\sigma_v$ is the stellar velocity dispersion, and we
start from a spherical unperturbed (early-type) galaxy with potential
$U_0$. In other words, in the unperturbed galaxy the DM
supplies the exact contribution to have the
given potential $U_0$ and the King distribution function for $f_\star$.

We take an
external tidal potential and add it to the (unperturbed) galaxy potential, thus
ignoring the changes in the galaxy potential induced by the deformation of the
galaxy (for a similar not self-consistent approach, see
\citealp{article--Ciotti_Dutta--1994--alignment}; see also \citealp{
article--Bertin_Varri--2008--construction}). As pointed out by
\cite{article--Ciotti_Dutta--1994--alignment}, it is possible to neglect the
actual self-force of the galaxy because it \emph{\textup{decreases}} from the center of the
galaxy (see, e.g.,\ the potential \eqref{eq:self-field} of the unperturbed
self-field), whereas the external tidal force \textup{increases}.
Therefore, we take $U=U_0+U_\mathrm{tidal}$ (see
\citealp{article--Bertin_Varri--2008--construction,
article--Varri_Bertin--2009--properties} for a self-consistent approach applied
to a globular stellar cluster).  We
take a weak external tidal field $|U_\mathrm{tidal}/\sigma_v^2|\ll1$ and expand the stellar
distribution function with respect to it
\begin{align}
        f_\star={}& A\mathrm e^{-U_0(\vec x)/\sigma_v^2-\|\vec{\dot
        x}\|^2/(2\sigma_v^2)}\big(1-\sigma_v^{-2}\Phi_{ij}x_ix_j+\ldots\big)-A\mathrm
        e^{-E_\mathrm{tr}/\sigma_v^2}\notag\\
        ={}& f_\star^{(0)}+f_\star^{(1)}+\ldots\;,\\
        f_\star^{(0)}={}&A\Big(\mathrm e^{-U_0(\vec x)/\sigma_v^2-\|\vec{\dot
        x}\|^2/(2\sigma_v^2)}-\mathrm
        e^{-E_\mathrm{tr}/\sigma_v^2}\Big)\;,\\
        f_\star^{(1)}={}&-\sigma_v^{-2}\Phi_{ij}x_ix_jA\mathrm e^{-U_0(\vec
        x)/\sigma_v^2-\|\vec{\dot x}\|^2/(2\sigma_v^2)}\;,
\end{align}
where $\Phi_{ij}=\frac12 \partial_i \partial_j
U_\mathrm{ext}$ is the \emph{\textup{tidal tensor}} and
$U_\mathrm{tidal}=\frac12x_ix_j\partial_i\partial_jU_\mathrm{ext}=\Phi_{ij}x_ix_j$
is the \textup{tidal field} (potential).

\begin{figure}
      \centering
      \resizebox{\columnwidth}{!}{\input{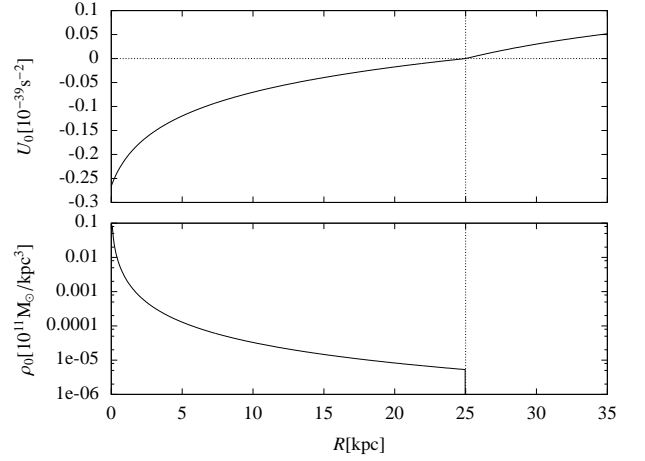}}
      \caption{Self-potential~\eqref{eq:self-field} and total mass
      distribution~\eqref{eq:rho-zero} of the unperturbed galaxy. The values
      used are $r_\mathrm{core}=\unit[1]{kpc}$,
      $r_\mathrm{tr}=\unit[25]{kpc}$ and $M=\unit[10^{11}]{M_\sun}$. The dashed
      vertical line indicates the truncation radius $r_\mathrm{tr}$.}
      \label{fig:U0}
\end{figure}

This expansion is valid only inside the galaxy
\citep{article--Bertin_Varri--2008--construction}; in fact, even if the exponential can always be expanded (when $|U_\mathrm{tidal}/\sigma_v| \ll 1$), this
expansion is not granted to be meaningful unless the perturbed self-potential
$U_0+U_\mathrm{tidal}$ is not too different from the actual galaxy self-potential, $U$. This condition breaks down near the galactic boundaries because, as
stated above, the galaxy self-potential decreases toward the boundaries, whereas
the external tidal field increases (since $\Phi_{ij}$ is traceless, the
isopotential surfaces may not even be ellipses when $|U_\mathrm{tidal}|\simeq
|U_0|$). From Eq.~\eqref{eq:luminous-quadrupole}, changing integration variables
to those of standard two-dimensional vectors, considering that $I(\vec x)\propto
\int\rho_\star(\vec x)\,\mathrm dx_3$, and using Eq.~\eqref{eq:poisson-f}, we
obtain for the luminous quadrupole
\begin{multline}
        \label{eq:quadrupole-expansion}
        \tens Q_{ij}=\frac1{D_\mathrm{s}^2N}\int
        \big(f_\star^{(0)}+f_\star^{(1)}+\ldots\big)x_ix_j\,\mathrm d^3\!x\,\mathrm d^3\!\dot x\\ =\tens
        Q_{ij}^{(0)}+\tens Q_{ij}^{(1)}+\ldots\;,
\end{multline}
where $D_\mathrm{s}$ is the angular-diameter distance to the
galaxy and $N$ is the number of stars of the galaxy. The integration of
Eq.~\eqref{eq:quadrupole-expansion} has to be carried out (i) in the region
\begin{equation}
        \label{eq:v-max}
        \|\vec{\dot x}\|\le v_\mathrm{max}(r)=\sqrt{2\big(E_\mathrm{tr}-U_0(r)\big)}
\end{equation}
because of the energy truncation in the distribution function and (ii) for
$\|\vec{x}\|\le r_\mathrm{max}<r_\mathrm{tr}$, a condition that mimics a real weak-lensing measurement, where the integration in the luminous quadrupole is carried
out based on a window function. The quantity $r_\mathrm{max}$ is chosen so as to
avoid regions where the external tidal field becomes similar
to or larger than
the galactic field, that is, the condition
$|U_\mathrm{tidal}(r_\mathrm{max})|\ll|U_0(r_\mathrm{max})|$ holds (see
discussion above). The zeroth
order of the expansion of the luminous quadrupole $\tens Q_{ij}^{(0)}$ is
proportional to the identity matrix because the unperturbed galaxy field is
spherical:
\begin{align}
        \label{eq:Q-0}
        \tens Q_{ij}^{(0)}={}&\frac{(4\pi)^2 A}{3ND_\mathrm{s}^2}\Big(\mathcal
        F_1-\mathcal F_2\Big)\,\delta_{ij}\;\\
        \label{eq:F1}
        \mathcal F_1={}&\int_0^{r_\mathrm{max}}\mathrm dr\,r^4\mathrm
        e^{-U_0(r)/\sigma_v^2}\int_0^{v_\mathrm{max}(r)}v^2\mathrm
        e^{-v^2/(2\sigma_v^2)}\,\mathrm dv\notag\\
        ={}&\sigma_v^2 \int_0^{r_\mathrm{max}}r^4 \mathrm
        e^{-U_0(r)/\sigma_v^2}
        \Bigg[-v_\mathrm{max}(r) \mathrm
        e^{-v_\mathrm{max}^2(r)/(2\sigma_v^2)}\notag\\
        &+\sigma_v\sqrt{\frac\pi2}\,
        \mathop{\mathrm{erf}}\left(\frac{v_\mathrm{max}(r)}{\sqrt2\sigma_v}\right)
        \Bigg]\,\mathrm dr\;,\\
        \label{eq:F2}
        \mathcal F_2={}&\frac{\mathrm
        e^{-E_\mathrm{tr}/\sigma_v^2}}3\int_0^{r_\mathrm{max}}r^4\Big(v_\mathrm{max}(r)\Big)^3\mathrm
        dr\;.
\end{align}
We use a galaxy logarithmic potential $U_0$ (Fig.~\ref{fig:U0}) to model the
total mass distribution in a galaxy (luminous plus dark matter;
\citealp[see e.g.{}][]{book--Bertin--2000--dynamics, paper--Koopmans_al--2006--sloan})
\begin{align}
        \label{eq:self-field}
        U_0(\vec
        x)={}&\begin{cases}\displaystyle MG\frac{r_\mathrm{tr}+r_\mathrm{core}}{r_\mathrm{tr}^2}\log\left(\frac{r_\mathrm{core}+\|\vec
        x\|}{r_\mathrm{core}+r_\mathrm{tr}}\right)&\mbox{if }\|\vec
        x\|<r_\mathrm{tr}\;,\\
        \displaystyle -MG\left(\frac1{\|\vec
        x\|}-\frac1{r_\mathrm{tr}}\right)&\mbox{if }\|\vec x\|>r_\mathrm{tr}\;,\end{cases}\\
        \label{eq:rho-zero}
        \rho_0(\vec
        x)={}&\begin{cases}\displaystyle \frac{M}{4\pi}\frac{r_\mathrm{tr}+r_\mathrm{core}}{r_\mathrm{tr}^2}\frac{2r_\mathrm{core}
        +\|\vec x\|} {\|\vec x\|\big(r_\mathrm{core} + \|\vec x\|\big)^2}&\mbox{if
        }\|\vec x\|<r_\mathrm{tr}\;,\\
        \displaystyle 0&\mbox{if }\|\vec x\|>r_\mathrm{tr}\;,
        \end{cases}
\end{align}
where $M=\unit[10^{11}]{M_\sun}$ is the galaxy mass, $U_0(r_\mathrm{tr})=0$,
$r_\mathrm{core}=\unit[1]{kpc}$ is the core radius, and
$r_\mathrm{tr}=\unit[25]{kpc}$ is the truncation radius. Throughout this paper we
take for an elliptical galaxy $\sigma_v\simeq\unit[200]{km\,s^{-1}}$ and
$r_\mathrm{max} = \unit[10]{kpc}$ (see discussion above). We then obtain
\begin{align}
        \label{eq:value-F1}
        \mathcal F_1&{}\simeq\unit[6.1\times10^{-5}]{Mpc^5\,(km/s)^3}\;,\\
        \label{eq:value-F2}
        \mathcal F_2&{}\simeq\unit[4.6\times10^{-5}]{Mpc^5\,(km/s)^3}\;.
\end{align}

The first-order perturbation to the luminous quadrupole is
\begin{equation}
\begin{split}
\tens Q_{ij}^{(1)}&{}=-\frac{A\Phi_{lm}}{ND_\mathrm{s}^2\sigma_v^2}\int x_l x_m x_i x_j
\mathrm e^{-U_0(\vec x)/\sigma_v^2-\|\vec{\dot
x}\|^2/(2\sigma_v^2)}\mathrm d^3\!x\,\mathrm
d^3\!\dot x\\
&{}=-\frac{(4\pi)^2 A\mathcal F_3}{15\sigma_v^2ND_\mathrm{s}^2}
\cdot\Big(2\Phi_{ij}+\delta_{ij}\mathop{\mathrm{Tr}}\Phi\Big)\;,
\end{split}
\end{equation}
where $\mathop{\mathrm{Tr}}\Phi=\Phi_{11}+\Phi_{22}+\Phi_{33}$, and
\begin{align}
        \label{eq:F3}
        \mathcal F_3={}&\sigma_v^2 \int_0^{r_\mathrm{max}}r^6 \mathrm
        e^{-U_0(r)/\sigma_v^2}
        \Bigg[-v_\mathrm{max}(r) \mathrm e^{-v_\mathrm{max}^2(r)/(2\sigma_v^2)}\notag\\
        &+\sigma_v\sqrt{\frac\pi2}\,
        \mathop{\mathrm{erf}}\left(\frac{v_\mathrm{max}(r)}{\sqrt2\sigma_v}\right)\Bigg]\,\mathrm dr\\
        \simeq{}&\unit[4.9\times10^{-9}]{Mpc^7\,(km/s)^{3}}\notag\;.
        \end{align}
From Eq. \eqref{eq:complex-ellipticity}, considering that $\tens
Q^{(0)}_{ij}\propto\delta_{ij}$ (spherical galaxy), we find for the complex ellipticity
\begin{equation}
\label{eq:eps-final}
\begin{split}
\varepsilon^\mathrm{s}\simeq{}&\frac{\tens Q^{(1)}_{11}-\tens Q^{(1)}_{22}+2\imath \tens
Q^{(1)}_{12}}{\mathop{\mathrm{Tr}}\tens Q^{(0)}+2\sqrt{\mathrm{det}\,\tens Q^{(0)}}}\\
={}&-\frac 1{10\sigma_v^2}\frac{\mathcal F_3}{\mathcal
F_1-\mathcal
F_2}\Big(\Phi_{11}-\Phi_{22}+2\imath\Phi_{12}\Big)\;.
\end{split}
\end{equation}
Finally, we obtain the general expression for the ellipticity induced on
a particular galaxy by an external tidal field. This expression can be
factorized in two parts: one part only depends on the particular galaxy we are
considering (its mass, its size, its self-field, and the velocity dispersion of
its stars); the other part only depends on the tidal field, with the
expected dependence
\citep[e.g.,][]{article--Catelan_Kamionkowski_Blandford--2001--intrinsic}. The
difference with
\citet{article--Catelan_Kamionkowski_Blandford--2001--intrinsic} is that in
Eq.~\eqref{eq:eps-final} the external field is that acting at the moment of the
light emission, and not that acting at the formation epoch.
Comparing Eq.~\eqref{eq:eps-final} with Eqs.~\eqref{eq:catelan1} and
\eqref{eq:catelan2}, we obtain the analytic expression for the constant of
proportionality appearing in Eqs.~\eqref{eq:catelan1} and \eqref{eq:catelan2}
\begin{equation}
\label{eq:C-prop}
C=-\frac{1}{20\sigma_v^2}\frac{\mathcal F_3}{\mathcal F_1-\mathcal
  F_2}\simeq\unit[-4.2\times10^{14}]{yr^2}\;.
\end{equation}
Equation~\eqref{eq:eps-final} is the main result of this paper: it
provides a direct way to estimate the intrinsic ellipticity of an
early-type galaxy that is subject to an external tidal field.

To better appreciate this result, consider a thin lens such as a
galaxy cluster.  In this case, we can write the lens shear as
\begin{equation}
\begin{split}
\gamma={}&c^{-2}\frac{D_\mathrm{ls}D_\mathrm{l}}{D_\mathrm{s}}
\Big(\partial_{11}-\partial_{22}+2\imath\partial_{12}\Big)\int
\mathrm dx_3 U(\vec x)\\
\simeq{}& c^{-2}\frac{D_\mathrm{ls}D_\mathrm{l}L_\mathrm{cl}}{D_\mathrm{s}}
\Big(\Phi_{11}-\Phi_{22}+2\imath\Phi_{12}\Big)\;,
\end{split}
\end{equation}
where the derivatives are given with respect to the physical coordinates, $D_\mathrm{s}$,
$D_\mathrm{l}$ and $D_\mathrm{ls}$ are the angular distance of a source galaxy,
of the lens, and between the lens and the source galaxy, respectively, and
$L_\mathrm{cl}$ is the galaxy cluster typical size.  Interestingly, the
expression for $\gamma$ has the same form as Eq.~\eqref{eq:eps-final}, that is,
it depends on the same combination of second-order partial derivatives of the
tidal tensor $\Phi$.  The constants involved, instead, are clearly different: in
particular, for a typical galaxy cluster with $L_\mathrm{cl}\simeq\unit[1]{Mpc}$
at $z_\mathrm{l} = 0.5$ and for a source at $z_\mathrm{s} = 1$ we find that the
coupling constant between the shear $\gamma$ and the tidal field is about $-20 C$.  As a
result, in this context (for a galaxy cluster) we expect that the intrinsic
alignment produced by the instantaneous tidal field is negligible.

\subsection{Particular cases}

\begin{figure}
      \centering
      \resizebox{\columnwidth}{!}{\input{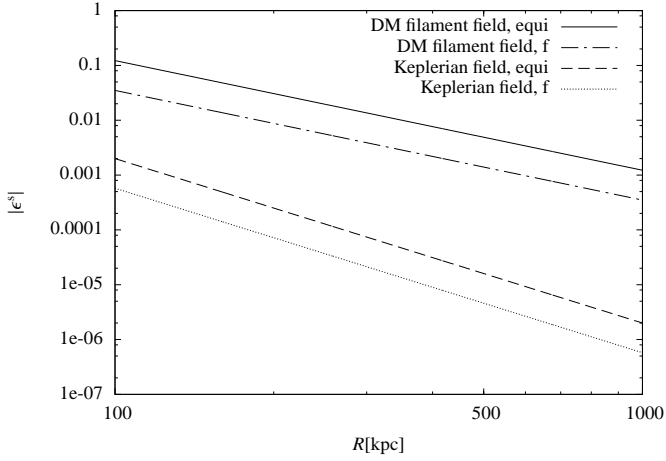}}
      \caption{Radial dependence of the intrinsic galaxy ellipticity in the
      particular cases (Keplerian and DM filament fields) considered in this
      paper, computed through the equipotential approximation (equi, see
      Appendix~\ref{app:isopotential-app}) and through the distribution
      function formalism (f, see Sect.~\ref{sec:dynamical-app}).  Note the
      different radial dependence of the Keplerian ($R^{-3}$) and the DM
      filament ($R^{-2}$) cases; the equipotential approximation and the
      distribution function formalism yield the same radial dependence in the
      two cases, but different multiplicative factors (see the text for
      details).}
      \label{fig:eps}
\end{figure}

We now consider an external Keplerian field, that is, the potential generated by a second
galaxy of mass $M_\mathrm{ext}$, assumed to be spherical and centered at $\vec R$. The tidal
acceleration and the tidal potential acting on a star of the first galaxy are
\begin{align}
        a_{i,\mathrm{tidal,}M}&{}=-\frac{GM_\mathrm{ext}}{R^3}\Bigg(\delta_{ij}-3\frac{R_iR_j}{R^2}\Bigg)x_j\;,\\
        \label{eq:tidal-kepler}
        U_{\mathrm{tidal,}M}&{}=+\frac{GM_\mathrm{ext}}{2R^3}\Bigg(\delta_{ij}-3\frac{R_iR_j}{R^2}\Bigg)x_ix_j\;.
\end{align}
If we place the external galaxy at $\vec R=(R,0,0)$, using Eq.~\eqref{eq:eps-final}, we
obtain (Fig.~\ref{fig:eps})
\begin{equation}
        \label{eq:eps-f-value}
                \varepsilon^\mathrm s=+\frac{3GM_\mathrm{ext}}{20\sigma_v^2R^3}\frac{\mathcal
                F_3}{\mathcal F_1-\mathcal
                F_2}\simeq+ 4.6\times10^{-6}\;,
\end{equation}
where in the last step we used $M_\mathrm{ext}=\unit[10^{11}]{M_\sun}$ and
$R=\unit[500]{kpc}$.  We note that $\varepsilon^\mathrm s=\mathop{\mathrm
O}(R^{-3})$.

We now examine a different case: the external field is generated by a dark
matter (DM) filament, and the galaxy is outside it. We are considering a
tidal field generated by an (infinite) linear distribution of matter along
the line of sight $\vec{\hat x_3}$
\begin{align}
        \label{eq:F-tidal-linear}
        a_{i,\mathrm{tidal,\lambda}}&{}=-\frac{2G\lambda
        }{R^2}\Bigg(\tilde\delta_{ij}-2\frac{R_iR_j}{R^2}\Bigg)x_j\;,\\
        \label{eq:U-tidal-linear}
        U_\mathrm{tidal,\lambda}&{}=+\frac{G\lambda}{R^2}\Bigg(\tilde\delta_{ij}-2\frac{R_iR_j}{R^2}\Bigg)x_ix_j\;,
\end{align}
where $\vec R$ is the distance of the galaxy from the DM straight line (the
filament), $\tilde\delta=\mathop{\mathrm{diag}}(1,1,0)$ and $\lambda$ is the
linear mass distribution of the dark matter filament. Obviously,
$a_{3\mathrm{,tidal,\lambda}}=0$.  We now need to estimate the quantity
$\lambda$, the linear density of the DM filament. DM filaments are thought to
extend between massive galaxy clusters
\citep[e.g.,][]{article--Springel_all--2005--simulations}. We therefore
consider the number density of massive clusters and estimate the average
distance $d_\mathrm{cl}$ between two clusters. To do so, we use the
cumulative distribution of galaxy clusters measured by
\cite{article--Bahcall_Cen--1993--mass}:
\begin{align}
        \label{eq:n-clusters}
        n(>M_\mathrm{cl})&{}=\unit[4\times10^{-5}\frac{M^\ast}{M_\mathrm{cl}}\mathrm
        e^{-\frac{M_\mathrm{cl}}{M^\ast}}h^3]{Mpc^{-3}}\;,\\
        d_\mathrm{cl}(M_\mathrm{cl})&{}=\Big(n(>M_\mathrm{cl})\Big)^{-1/3}\;,
\end{align}
where $M^\ast=\unit[(1.8\pm0.3)\times10^{14}h^{-1}]{M_\sun}$ and $h=0.7$.
Setting $M_\mathrm{cl}=M^\ast$ leads\footnote{An important point to keep in mind is that the
large-scale structure is strongly time-dependent and that the mean
high-mass-cluster distance changes with time
\citep{article--Springel_all--2005--simulations}. Here, we are considering
the local Universe, that is, regions where the redshift is much lower than unity.} to
\begin{equation}
d_\mathrm{cl}\simeq\unit[58]{Mpc}\; ,
\end{equation}
which is consistent with the typical value reported by
\cite{article--Springel_all--2005--simulations}, that is, $\unit[100]{Mpc}$.
We assume
that $34\%$ of the mass in the Universe is concentrated in filaments, as the
numerical results of \cite{paper--Hoffman_al--2012--kinematic} indicates. We then have that
\begin{equation}
        \label{eq:lambda}
        \lambda(M_\mathrm{cl})\simeq0.34\rho_{\mathrm m}
        d^2_\mathrm{cl}\simeq\unit[4.67\times10^7]{M_\sun\,pc^{-1}}\;.
\end{equation}
If we place
the filament in such a way
that it crosses the source plane at $\vec R=(R,0)$, with $R=\unit[500]{kpc}$, and use
Eq.~\eqref{eq:eps-final}, we obtain (Fig.~\ref{fig:eps})
\begin{equation}
        \label{eq:eps-f-linear}
                \varepsilon^\mathrm s=+\frac{G\lambda}{5\sigma_v^2R^2}\frac{\mathcal
                F_3}{\mathcal F_1-\mathcal F_2} \simeq+1.4\times10^{-3}\;.
\end{equation}
We observe that in this case the dependence on the distance is $\varepsilon^\mathrm
s=\mathop{\mathrm O}(R^{-2})$.

\section{Discussion}
\label{sec:discussion}
\subsection{Comparison with literature}
It is common \citep[e.g.,][]{paper--Joachimi_Bridle--2010--simultaneous,
paper--Joachimi_al--2011--constraints} to consider the galaxy number
density contrast-intrinsic ellipticity correlation, that is,
\begin{equation}
  \label{eq:C_nI}
  C_{\mathrm{gI}}(x)=\langle \delta_n(\mathbf x_0)\varepsilon^{\mathrm s}_+(\mathbf x_0
  +\mathbf x)\rangle_{x_0} \; ,
\end{equation}
where the correlation function only depends on the modulus of $\mathbf x$
because isotropy, $\delta_n$ is the galaxy number density contrast, and $\varepsilon^{\mathrm s}_+$ is the
real part of the
complex (intrinsic) ellipticity in a coordinate system with the real axis along the line
connecting the two correlating points $\mathbf x_0$ and
$\mathbf x$, as in Eqs.~\eqref{eq:C1}, \eqref{eq:C2}, and \eqref{eq:C12}. In contrast to Eqs.~\eqref{eq:C1} and following, the correlation in
Eq.~\eqref{eq:C_nI} is
carried on in the three-dimensional real space.
We now define the Fourier transform $\tilde f(\mathbf k)$ of a function $f(\mathbf x)$ as
\begin{equation}
        \tilde f(\mathbf k) = \int f(\mathbf x) \mathrm e^{\imath \mathbf k \cdot \mathbf
        x}\mathrm d^3\!x\;.
\end{equation}
Since
\begin{align}
        \tilde {\delta_n}(\mathbf k) ={}& b_{\mathrm g}\tilde \delta(\mathbf k)\;,\\
        \label{eq:tilde_eps+}
        \tilde \varepsilon^{\mathrm s}_+(\mathbf k) ={}& -C (k_1^2-k_2^2) \tilde
        U(\mathbf k)= 4 \pi
        C G \frac{k_1^2-k_2^2}{k^2}\rho_{\mathrm{cr}}\Omega_{\mathrm m}\tilde
   \delta(\mathbf k)\;,
\end{align}
where $b_{\mathrm g}$ is the galaxy bias and in the second equality of
Eq.~\eqref{eq:tilde_eps+} we used Eq.~\eqref{eq:catelan1}, the Poisson
equation and the relation $\tilde
\rho(\mathbf k)=\rho_{\mathrm{cr}}\Omega_{\mathrm m}\tilde \delta(\mathbf k)$ (we
neglect the three-dimensional Dirac delta in the origin). If we
introduce the elliptical galaxy fraction $f_{\mathrm{ell}}$ (because
only elliptical galaxies align in our model), we find
\begin{multline}
        \label{eq:P_gI}
        (2\pi)^3\delta^{(3)}(\mathbf k-\mathbf{k'})P_{\mathrm {gI}}(k,z)=\langle \tilde
        {\delta_n}(\mathbf k)\tilde\varepsilon^{\mathrm s}_+(\mathbf{k'})\rangle\\
        \simeq (2\pi)^3\delta^{(3)}(\mathbf k -
        \mathbf{k'})4\pi b_{\mathrm g}f_{\mathrm{ell}} C
        G\rho_{\mathrm{cr}}\Omega_{\mathrm m} P_\delta(k,z)\;,
\end{multline}
where $P_\delta(k,z)$ is the density contrast power spectrum at the
galaxy redshift $z$ and where we have approximated the term $(k_1^2 -
k_2^2) / k^2$ to unity.  We may now consider Eqs.~(6) and (19) of
\citet{paper--Joachimi_al--2011--constraints}, that is,
\begin{equation}
        \label{eq:joachimi+2011}
        P^{\mathrm{(JMAB)}}_{\mathrm{gI}}(k,z,L)=-AC_1b_{\mathrm g}\rho_{\mathrm{cr}}
        \frac{\Omega_{\mathrm m}}{D(z)}P_\delta(k,z)
        \left(\frac{1+z}{1+z_0}\right)^\eta \left(\frac{L}{L_0}\right)^\beta,
\end{equation}
where $D(z)$ is the linear growth factor normalized to unity today, $P_\delta$
is the (nonlinear) power spectrum of the density contrast, $z$ and $L$ are the redshift and the
absolute luminosity of the early-type galaxy, and the arbitrary reference values
are $z_0=0.3$ and $L_0$, the latter corresponding to an absolute r-band
magnitude of $-22$, passively evolved to $z=0$. The parameters $AC_1$, $\eta,$ and
$\beta$ have been measured by \citet{paper--Joachimi_al--2011--constraints} for
comoving transverse separations greater than $\unit[6h^{-1}]{Mpc}$,
\begin{align}
        \label{eq:C_joachimi}
        AC_1\rho_{\mathrm{cr}}&{}=0.077\pm0.008\;,\\
        \beta&{}=1.13^{+0.25}_{-0.20}\;,\\
        \eta&{}=-0.27^{+0.80}_{-0.79}\;.
\end{align}
Comparing Eq.~\eqref{eq:P_gI} to Eq.~\eqref{eq:joachimi+2011} and neglecting the
redshift and luminosity dependence, we obtain that Eq.~\eqref{eq:C_joachimi}
corresponds to the term
\begin{equation}
        \label{eq:our-const}
        -4\pi CG\rho_{\mathrm{cr}}\simeq3.2\times10^{-6}
\end{equation}
of Eq.~\eqref{eq:P_gI}, where we set $f_{\mathrm{ell}}=1$ because
\citet{paper--Joachimi_al--2011--constraints} considered only
early-type galaxies for the shape measurements. The result is clearly
inconsistent. It is unlikely that this inconsistency is due to our use
of the simple King model because the different approach adopted in
Appendix~\ref{app:isopotential-app}, where we do not make use of the
stellar distribution function, yields similar results
(cf. Eqs.~\eqref{eq:eps-f-value}, \eqref{eq:eps-f-linear},
\eqref{eq:epsilon-iso-kepler} and \eqref{eq:eps-iso-linear}). Then, at
least at large scales ($r_p>\unit[6h^{-1}]{Mpc}$), our model does not
reproduce the observations.

The incompatibility of Eq.~\eqref{eq:our-const} with Eq.~\eqref{eq:C_joachimi}
means that the galaxy deformation due to the instantaneous external tidal field
cannot yield the observed IA signal. A possible
explanation is that the IA signal is caused by the galaxy formation process
and/or its merging history. To obtain analytic results, these processes therefore need to be linked to the external tidal field. In particular, at least
for the merging history, the velocity shear field needs to be considered because it
was recently discovered \citep{paper--Hoffman_al--2012--kinematic,
paper--Libeskind_al--2014--universal, paper--Lee_Choi--2014--detection} that
mergers preferentially occur along the velocity shear minor eigenvector. A
detailed analysis of this process is beyond the aims of this paper.

\subsection{Halo case}

Until now, we have only considered the galaxy with its total (luminous plus dark)
mass. We have stopped the integration of the luminous quadrupole at
$r_\mathrm{max}<r_\mathrm{tr}$. However, the
ellipticity of the whole DM halo remains unclear. We remark that it is not obvious that
Eq.~\eqref{eq:complex-ellipticity} is the correct choice for the halo ellipticity because the halo luminous quadrupole cannot
be measured in real surveys. For the
same reason, it is unclear how to choose $r_\mathrm{max}$, but for the
condition $|U_0(r_\mathrm{max})|\gg |U_\mathrm{tidal}(r_\mathrm{max})|$.
Still, it is interesting to consider the extreme case of a dark matter halo
$M=\unit[2\times10^{13}]{M_\sun}$, $r_\mathrm{tr}=\unit[560]{kpc}$ and
$\sigma_v=\unit[250]{km\,s^{-1}}$, as found by
\cite{paper--Gavazzi_al--2007--sloan} from an
ensemble average over 22 \emph{\textup{massive}} halos. Even now, we obtain (taking $r_\mathrm{core} =
0.01\cdot r_\mathrm{tr}$ and $r_\mathrm{max}=0.80\cdot r_\mathrm{tr}$)
\begin{align}
\label{eq:C-halo-value}
C_\mathrm{halo}&{}\simeq\unit[-1.3\times10^{17}]{yr^2}\;,\\
\label{eq:halo-value}
-4\pi C_\mathrm{halo}G\rho_\mathrm{cr}&{}=0.0042\;,
\end{align}
which is still almost a factor $\sim 20$ lower than the value of
Eq.~\eqref{eq:C_joachimi}, that is, even if instantaneous tidal fields produce a greater
IA on dark matter halos than on galaxies, the effect is still much too weak to
account for the observed signal. Arguably, the galaxies observed by
\citet{paper--Joachimi_al--2011--constraints} are, on average, less massive objects, and therefore the
value of Eq.~\eqref{eq:C-halo-value} should be taken as an upper bound. In addition,
we stopped the integration near the halo boundaries
($r_\mathrm{max}=0.80\cdot r_\mathrm{tr}$), and the resulting ellipticity is
therefore greater than that we would obtain with a smaller
$r_\mathrm{max}$. This
shows that our model is essentially inadequate to explain the observed amount of
IA.

\section{Conclusions}
\label{sec:conclusions}

The exact origin of intrinsic alignment of galaxies is still unclear, and there are no analytic estimates
of the amount of IA caused by different possible processes.
We here support the standard view on IA (i.e., it is caused by
formation and accretion processes) by a \emph{\textup{reduction ad absurdum}}. To arrive at this result, we estimated the amount of IA in elliptical galaxies due to an
external instantaneous tidal field by considering the galaxy stellar distribution
function and the luminous quadrupole. In addition, in
Appendix~\ref{app:rigid-body}, we determined the typical oscillation time-scale
for an elliptical galaxy modeled as a rigid body, subject to an external tidal
field, and in Appendix~\ref{app:isopotential-app} we determined the ellipticity
of a galaxy (subject to an external instantaneous tidal field) in a different
way by studying its equipotential surfaces. The main results are the following:
\begin{enumerate}
\item The distribution function approach allows us to
  analytically determine the dependence of the intrinsic ellipticity
  on the tidal field, heuristically formulated by
  \cite{article--Catelan_Kamionkowski_Blandford--2001--intrinsic}, in
  terms of the properties of the galaxy (i.e., its mass, size,
  velocity dispersion, and stellar distribution function),
  Eqs.~\eqref{eq:eps-final} and \eqref{eq:C-prop}.\item The intrinsic alignment signal obtained when our model is
  applied to an elliptical galaxy is negligible with respect to the
  observed one (at least at large scales, $>\unit[6h^{-1}]{Mpc}$,
  \citealp{paper--Joachimi_al--2011--constraints}), cf.{}
  Eqs.~\eqref{eq:C_joachimi} and \eqref{eq:our-const}. Thus one has to
  consider the galaxy formation process and/or its merging history.
\item When our model is applied to the
  whole galaxy halo, the intrinsic alignment signal increases (but it is still
  inconsistent with the observed signal), Eq.~\eqref{eq:halo-value}.
\end{enumerate}

The work we described here is a step toward a simple physical
understanding of the bias introduced in weak-lensing measurements by IA.  In the
past, IA has been regarded as a source of concern for cosmic-shear measurements.
More recently, it has been regarded as an opportunity to investigate the
physical properties of galaxies, their DM halos, and their formation history.
In this perspective, it is important to develop analytic models of IA, such as
those presented here, to be able to interpret the results of future weak-lensing surveys.  In this respect, a complete model of IA could even be used
directly to reconstruct the local tidal field acting on elliptical galaxies.

The theoretical analysis could be improved by applying the techniques
we presented (i.e., the stellar distribution function method)
to the study of the formation and accretion processes, in order to
determine their contributes to the IA signal. It would also be interesting to study the contribution to IA due to \emph{\textup{continuously}} applied
tidal fields (instead of only considering the tidal field acting at the
emission epoch, as we have did here), and the evolution of
IA with redshift.

\begin{acknowledgements}
We wish to thank Giuseppe Bertin and Benjamin Joachimi for invaluable
suggestions and very useful discussions, which significantly improved the paper.
We also thank the anonymous referee for his or her helpful suggestions concerning the
presentation of this paper.
G.~C. thanks Malegori's family ("Franca Erba" scholarship) and BCC Carugate e
Inzago Bank ("Gildo Vinco" award) for supporting his studies.
\end{acknowledgements}

\bibliographystyle{aa}
\bibliography{bibliografia}

\appendix

\section{Rigid-body approximation}
\label{app:rigid-body}

In this appendix we compare the time scale of motions of galaxies modeled as
rigid bodies with the time scale of their internal dynamics.  In particular, we
consider the (rigid) oscillations of elliptical galaxies and the precession of
spiral galaxies. We show below that internal dynamics is much faster than rigid-body dynamics, thus confirming the results of
\cite{article--Ciotti_Dutta--1994--alignment} for elliptical galaxies.

\subsection{General case}

We assume that the distribution of mass of the galaxy
only depends on the elliptical radius $\tilde r$ and that its principal axes
are aligned with the coordinate axes:
\begin{align}
  \rho(\mathbf x, \mathbf{\dot{x}}, t) = {} & \rho(\mathbf x) =
  \tilde \rho(\tilde r) \; , \\ 
  \tilde r = {} & \sqrt{\left(\frac {x_1} {r_1}\right)^2+\left(\frac
    {x_2}{r_2}\right)^2+\left(\frac {x_3}{r_3}\right)^2} \; ,
\end{align}
where $r_1$, $r_2$ and $r_3$ are the semi-axes of the ellipsoid ($\rho=0$ for
$\tilde r> 1$).
Then the inertia tensor is
\begin{equation}
  \tens I_{ij}=\int \rho(\mathbf x)\Big(\|\mathbf
  x\|^2\delta_{ij} - x_ix_j\Big) \, \mathrm d^3\!x \; ,
\end{equation}
where $\delta$ is the Kronecker delta, and we placed the center
of mass of the galaxy in the origin.  With this
definition, the moment of inertia $I_{\vec{\hat n}}$ along the unit
vector $\vec{\hat n}$ is
\begin{equation}
  \label{eq:inerzia-versore}
  \tens I_{\mathbf{\hat n}}=\hat n^i \tens I_{ij} \hat n^j\;.
\end{equation}
For an elliptical mass distribution whose principal axes are parallel
to the coordinate axes we obtain
\begin{align}
  \tens I= {} & \mathcal F_0
  r_1r_2r_3\mathop{\mathrm{diag}}\Big(r_2^2+r_3^2,r_1^2+r_3^2,r_1^2+r_2^2\Big) \; ,\\ 
  \mathcal F_0={}&\frac{4\pi}3 \int_0^1 \tilde\rho(\tilde r)\tilde
  r^4\,\mathrm d\tilde r \; .
\end{align}
The potential energy $W$ of a body subject to the action of the external field
$U_{\mathrm{ext}}(\mathbf x)$ (whose Laplacian is null in the
region occupied by the galaxy) is
\citep{book--Jackson--1998--classical} 
\begin{align}
  W= {} & \int\rho(\mathbf x)U_\mathrm{ext}(\mathbf x)\,\mathrm d^3\!x  \\
  {} \simeq {} & MU_\mathrm{ext}(\vec 0)+\frac16\mathcal
  Q_{ij}\partial^i\partial^jU_\mathrm{ext}(\vec 0)\; , \\ 
  \mathcal Q_{ij} = {} &\int \rho(\mathbf
  x)\Big(3x_ix_j-\|\mathbf x\|^2\delta_{ij}\Big)\, \mathrm d^3\!x \; , 
\end{align}
where $M$ and $\mathcal Q_{ij}$ are the mass and the quadrupole tensor of the
body.  For an elliptical mass distribution with principal axes parallel to the
coordinate axes we obtain
\begin{multline}
    \mathcal Q= \mathcal F_0
     r_1r_2r_3\mathop{\mathrm{diag}}\Big(2r_1^2-r_2^2-r_3^2,\\
     -r_1^2+2r_2^2-r_3^2,-r_1^2-r_2^2+2r_3^2\Big)\;.
\end{multline}

\subsection{Elliptical galaxies: oscillation period}

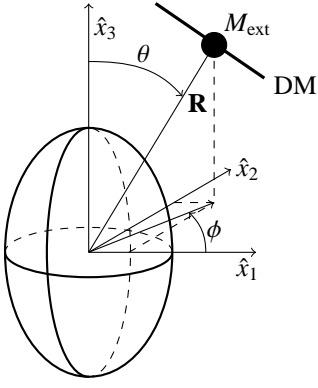
\begin{figure}
   \centering
   \begin{tikzpicture}[scale=1.1]
      \begin{scope}
         \path [clip] (0,-1.6) rectangle (-0.6,1.6);
         \draw [thick] (0,0) ellipse [x radius=.5,y radius=1.5];
      \end{scope}
      \begin{scope}
         \path [clip] (0,-1.5) rectangle (+0.5,1.5);
         \draw [dashed] (0,0) ellipse [x radius=.5,y radius=1.5];
      \end{scope}
      \draw [thick] (0,0) ellipse [x radius=1,y radius=1.5];
      \begin{scope}
         \path [clip] (-1,0) rectangle (1,-1.5);
         \draw [thick] (0,0) ellipse [x radius=1,y radius=0.3];
      \end{scope}
      \begin{scope}
         \path [clip] (-1,0) rectangle (1,1.5);
         \draw [dashed] (0,0) ellipse [x radius=1,y radius=0.3];
      \end{scope}
      \draw [->] (0,0) -- (0,3);
      \node at (.2,2.7) {$\hat x_3$};
      \draw [->] (0,0) -- (2,0);
      \node at (1.9,-0.2) {$\hat x_1$};
      \draw [->] (0,0) -- (1.7,1);
      \node at (1.9,1) {$\hat x_2$};
      \draw [fill] (1.5,2.5) circle [radius=0.15];
      \draw [->] (0,0) -- (1.5,2.5);
      \draw [->] (0,2.3) to [out=0,in=130] (1.12,1.88);
      \node at (.65,2.4) {$\theta$};
      \draw [dashed] (1.5,2.5) -- (1.5,0.6);
      \draw [->] (0,0) -- (1.5,0.6);
      \draw [dashed] (1.02,0.6) -- (1.5,0.6);
      \draw [dashed] (0.48,0) -- (1.5,0.6);
      \draw [->] (1.4,0) to [out=90,in=-45] (1.2,0.48);
      \node at (1.3,1.8) {$\vec R$};
      \node at (1.5,0.25) {$\phi$};
      \draw [very thick] (0.8,3) -- (2.1,2.1);
      \node [above right] at (1.5,2.5) {$M_\mathrm{ext}$};
      \node [above right] at (2.1,1.8) {DM};
   \end{tikzpicture}
   \caption{Spherical coordinate system adopted in the text, Keplerian
   ($M_\mathrm{ext}$) and DM filament cases.}
   \label{fig:rigid-body}
\end{figure}

For a Keplerian potential, the galaxy energy is
\begin{equation}
W_M=-\frac{GM_\mathrm{ext}M}{R}-\frac{GM_\mathrm{ext}\mathcal Q_{ij}}{6R^3}
\Bigg(3\frac{R_iR_j}{R^2}-\delta_{ij}\Bigg) \; .
\end{equation}
Without loss of generality, let $\mathcal Q$ be diagonal and let
$\theta$ and $\varphi$ be the angles between the $\vec{\hat x_3}$ axis
and $\vec R$ and between the $\vec{\hat x_1}$ axis and $\vec R -\vec
R\cdot\mathbf{\hat x_3}$, in a spherical coordinate system (see
Fig.~\ref{fig:rigid-body}).  In other words, we consider a coordinate system
whose axes are parallel to the principal axes of the galaxy.
Having kept $\mathcal Q$ diagonal, we have to rotate the tidal tensor.
Considering that $\mathop{\mathrm{Tr}}\mathcal Q=0$, we find
\begin{multline}
  \label{eq:energia}
        W_M =-\frac{GM_\mathrm{ext}M}{R}-\frac{GM_\mathrm{ext}}{2R^3}
        \Big(\mathcal Q_{11}\cos^2(\varphi)\sin^2(\theta) \\
         + \mathcal Q_{22}\sin^2(\varphi)\sin^2(\theta)+\mathcal
        Q_{33}\cos^2(\theta)\Big)\;.
\end{multline}
We can in principle determine the oscillation equations of the galaxy
via the Euler-Lagrange equations, using the correct kinetic energy
expression.  This approach leads to complex equations and
therefore we assume that (i) the elliptical distribution of mass is prolate
(i.e.,\ $r_1=r_2=r_\mathrm{eq}<r_3$), from which
$\tens I_{11}=\tens I_{22}=\tens I_\mathrm{eq}$ and $\mathcal
Q_{11}=\mathcal Q_{22}=\mathcal Q_\mathrm{eq}$, and (ii) there are no
initial intrinsic rotation ($\dot \varphi(0)=0$) or precession motions;
there could be only nutation ($\dot \theta(t)\ne 0$). Using the
Euler-Lagrange equations, we then obtain
\begin{equation}
\label{eq:oscillazione-theta}
\tens I_\mathrm{eq}\ddot\theta=\frac{GM_\mathrm{ext}}{2R^3}
\sin(2\theta)\cdot\big(\mathcal Q_\mathrm{eq}-\mathcal Q_{33}\big)\;.
\end{equation}
We explicitly note that this equation is not self-consistent: in fact,
in principle it should be completed with the equations of the relative
motion between the galaxy center of mass and the monopole.  However,
we are not interested in a complete treatment of the problem, but only
in an analytic estimate of the time scale of an oscillation.
Therefore we assume, without clarifying the physical mechanism
that would permit this, that the relative position between the external
monopole and the galaxy is fixed, and that only the orientation of the
galaxy can vary.  With these assumptions, Eq.
\eqref{eq:oscillazione-theta} is sufficient to describe the dynamics
of the system. To obtain the time scale of an oscillation, we
make the small-angle approximation ($\theta\ll1$) and recall that $\mathcal Q_{33}>
\mathcal Q_\mathrm{eq}$ (prolate galaxy), obtaining the pendulum equation.
Considering that the components of $\tens I$ and $\mathcal Q$
are of the same order of magnitude, we find
that the period $t_{\mathrm{osc,}M}$ of an oscillation is
\begin{align}
\label{eq:period-oscillation}
t_{\mathrm{osc,}M}&{}\simeq2\pi\sqrt{\frac{R^3}{GM_\mathrm{ext}}}\notag\\
&{}\simeq{}3.0\times10^{11}\times\Bigg(\frac{R}{\unit[1]{Mpc}}
\Bigg)^{\frac32}
\times\Bigg(\frac{M_\mathrm{ext}}{10^{11}M_\sun}\Bigg)^{-\frac12}
\unit{yr}\;.
\end{align}
We remark that Eq.~\eqref{eq:period-oscillation} is reminiscent of the expression of
the free-fall time $t_\mathrm{ff}\propto(G\rho)^{-1/2}\propto t_{\mathrm{osc,}M}
\times (D/2R)^{3/2}$, being $D$ the diameter of the galaxy.
Using the values adopted here for the external mass
$M_\mathrm{ext}=\unit[10^{11}]{M_\sun}$ and
for the distance $R=\unit[500]{kpc}$,
the period is
$t_{\mathrm{osc,}M}\simeq\unit[1.0\times10^{11}]{yr}$,
which is greater than the age of the Universe\footnote{It is interesting to consider the case of stellar globular clusters: assuming a distance
from the host galaxy of $R=\unit[30]{kpc}$,
we obtain a period $t_{\mathrm{osc,cl}}\simeq\unit[1.5\times10^9]{yr}$.
We emphasize that this value has to be increased by the multiplicative
factor $\sqrt{\tens I_\mathrm{eq}/(\mathcal Q_{33}-\mathcal Q_\mathrm{eq})}$.}
(the numerical integration of Eq.~\eqref{eq:oscillazione-theta} is in
accordance with the approximate result).

This period has to be compared with the time scale on which the shape of a
galaxy changes, because of the external gravitational field.
A rough estimate of this is the time necessary for a star to travel
across the galaxy \citep[e.g.,][]{article--Fleck_Kuhn--2003--parametric}
\begin{equation}
t_\mathrm{cross,ell}=\frac{D}{\sigma_v}\;,
\end{equation}
where $\sigma_v$ is the stellar velocity dispersion and $D$ is the diameter of the
galaxy. For $D\simeq\unit[40]{kpc}$ and $\sigma_v\simeq\unit[200]{km\ s^{-1}}$, we obtain
$t_\mathrm{cross,ell}\simeq\unit[2\times10^8]{yr}\ll t_{\mathrm{osc,}M}$.  In other words, an
elliptical galaxy, not aligned with the gravitational tidal field, deforms itself
before completing a rigid oscillation.

If we consider a DM filament along the $\vec{\hat x_1}$ axis, using
Eq.~\eqref{eq:U-tidal-linear},
we can write the galaxy energy
as
\begin{equation}
        \label{eq:W-tidal-linear}
                W_\lambda=+2MG\lambda\log\Bigg(\frac
                R{\ell'}\Bigg)-\frac{G\lambda\mathcal
                Q_{ij}}{3R^2}\Bigg(2\frac{R_iR_j}{R^2}-\tilde\delta_{ij}\Bigg)\;,
\end{equation}
where $\ell'$ is an arbitrary constant introduced to have an
adimensional logarithmic argument, and
$\tilde\delta=\mathop{\mathrm{diag}}(0,1,1)$.  As before, we rotate the coordinate
frame in such a way that $\mathcal Q_{ij}$ be diagonal\footnote{This
  procedure is more complicated than in the Keplerian case because
  $\tilde\delta$ is not a three-dimensional delta. It can be written
  as $\tilde\delta_{ij}=\delta_{ij}-\hat x_{i,1}\hat x_{j,1}$, where $\vec{\hat
    x_1}=(1,0,0)$ is the versor of the first axis, parallel to the DM
  filament. When we change the coordinate frame, we have to rotate $\vec
  R$ and $\vec{\hat x_1}$; we then obtain
  $W_\lambda(\theta,\phi)=\mathrm{const}-3^{-1}G\lambda
  R^{-2}(\mathcal Q_{11}\cos^2\phi(1+\sin^2\theta) +\mathcal
  Q_{22}\sin^2\phi(1+\sin^2\theta) + \mathcal
  Q_{33}(1+\cos^2\theta))\;$.} (see Fig.~\ref{fig:rigid-body}), we
use spherical coordinates and assume that the galaxy is prolate,
that there are no proper rotation nor precession motions, and that the
position of the galaxy is fixed. We then obtain
\begin{equation}
        \label{eq:lambda-oscillations-theta}
        \tens I_\mathrm{eq}\ddot\theta=\frac{G\lambda}{3R^2}\sin(2\theta)\cdot\big(\mathcal
        Q_\mathrm{eq}-\mathcal Q_{33}\big)\;.
\end{equation}
Using the same approximations as before and Eq.~\eqref{eq:lambda}, we may estimate the period of a small
rigid oscillation of the galaxy
\begin{align}
        t_\mathrm{osc,\lambda}&{}\simeq2\pi R\sqrt{\frac{3}{2G\lambda}}\notag\\
        \label{eq:t-osc-lambda}
        &{}\simeq1.0\times10^{10}\times\Bigg(\frac{R}{\unit[1]{Mpc}}\Bigg)\;\unit{yr}\;,
\end{align}
which in the case considered throughout the paper ($R=\unit[500]{kpc}$) corresponds to
$\unit[5.0\times10^9]{yr}$. This is shorter than the period of
oscillation found in the Keplerian case, but nevertheless greater
than $t_\mathrm{cross,ell}$. We therefore have to drop the
rigid-body approximation and deepen the description of elliptical galaxies to account for the internal degrees of freedom, as done in
Sect.~\ref{sec:dynamical-app} and Appendix~\ref{app:isopotential-app}.

\subsection{Spiral galaxies: precession period}

The key feature that distinguishes spiral galaxies from elliptical
ones is the dominance of ordered motions over chaotic ones, that is, the
characterizing presence of the angular momentum.  If we place a rigid body with an
angular momentum in an external field, it starts precessing.  In this
subsection we estimate the precession period for a spiral galaxy,
taken as a rigid body.

The precession period $t_\mathrm{prec}$ of a rotating rigid body is
\begin{equation}
        t_\mathrm{prec}=\frac{2\pi L\sin(\theta)}{\tau(\theta)}\;,
\end{equation}
where $L$ is the angular momentum of the spiral galaxy, $\theta$ is the angle
between $L$ and the external force, and $\tau(\theta)$ is the amount of the
momentum of the external force. In spiral galaxies the \emph{\textup{luminous}} mass
distribution is very different from that of the \emph{\textup{dark matter}} ; in our calculation we can detach
the two contributions because the DM distribution is (approximately)
spherical and does not generate a torque on the visible mass, in which we are
interested. To determine the total stellar angular momentum $L_\star$, we can use the rigid-body formula $L_\star=\tens I_{\star33}\omega_\star$. In spiral galaxies stars
at different distances
from the center of the galaxy have different angular velocities
($\omega_\star=\omega_\star(\ell)$,
where $\ell$ is the distance from the symmetry axis of the
spiral). Therefore it is sensible to use
a weighted angular velocity $\bar \omega_\star$, defined as
\begin{align}
        \bar \omega_\star & {} = L_\star / \tens I_{\star33} \notag\\
        & {} = \int\ell^2 v_\star(\ell)
        \Sigma_\star(\ell)\,\mathrm d\ell  \Biggm/ \int\ell^3
        \Sigma_\star(\ell)\,\mathrm d\ell\;,
\end{align}
where $\Sigma_\star(\ell)$ is the projected stellar surface density at
distance $\ell$ from the spiral symmetry axis and $v_\star(\ell)$ is
the stellar tangential velocity. In spiral galaxies, the stellar tangential
velocity is approximately constant, $v_\star(\ell)\simeq v_\star$, and
the stellar column density follows an exponential law with length scale
$\ell_0$, $\Sigma_\star(\ell)=\Sigma_0\exp(-\ell/\ell_0)$. We then obtain
\begin{equation}
        \label{eq:bar-omega}
        \bar\omega_\star=\frac{v_\star\int_0^\infty\ell^2\exp(-\ell/\ell_0) \,\mathrm d\ell}
        {\int_0^\infty\ell^3\exp(-\ell/\ell_0)\,\mathrm d\ell}=
        \frac{v_\star}{3\ell_0}\;.
\end{equation}
Using this expression in Eqs.~\eqref{eq:oscillazione-theta} and
\eqref{eq:lambda-oscillations-theta}, we find for an oblate galaxy
($r_3<r_1=r_2=r_\mathrm{eq}$)
\begin{align}
        t_{\mathrm{prec,}M}={}&\frac{2\pi v_\star}{3G\ell_0}\frac{\tens
        I_{\star33}}{\mathcal Q_{\star\mathrm{eq}}-\mathcal
        Q_{\star33}}\frac{R^3}{M_\mathrm{ext}}\frac1{\cos(\theta)}\;,\\
        t_\mathrm{prec,\lambda}={}&\frac{\pi v_\star}{G\ell_0}\frac{\tens
        I_{\star33}}{\mathcal Q_{\star\mathrm{eq}}-\mathcal
        Q_{\star33}}\frac{R^2}\lambda\frac1{\cos(\theta)}\;.
\end{align}
Assuming $v_\star=\unit[200]{km\,s^{-1}}$ and
$\ell_0=\unit[10]{kpc}$, we obtain
\begin{align}
        t_{\mathrm{prec,}M}={}&\unit[1.2\times10^{13}\frac{\tens I_{33}}{\mathcal
        Q_{11}-\mathcal Q_{33}}\Big(\cos(\theta)\Big)^{-1}]{yr}\;,\\
        t_\mathrm{prec,\lambda}={}&\unit[2.7\times10^{10}\frac{\tens I_{33}}{\mathcal
        Q_{11}-\mathcal Q_{33}}\Big(\cos(\theta)\Big)^{-1}]{yr}\;.
\end{align}
Therefore the time scale is longer than or similar to the age of the Universe,
and it is also longer than the deformation time of spiral galaxies
(similarly to that of the elliptical galaxy).

\section{Equipotential approximation}
\label{app:isopotential-app}

In Sect.~\ref{sec:dynamical-app} we have obtained the expression of the
intrinsic ellipticity of an early-type galaxy subjected to an external tidal
field. To do so, we have calculated the luminous quadrupole, making use of the
stellar distribution function. In this appendix we present another approach, which
is less complete but has the advantage of having a clear and simple physical
understanding. In particular, we model the deformation of the galaxy by means of the
equipotential surfaces of the total gravitational potential. In this approach,
we ``assume[s] that the local galaxy density is produced approximately by stars
near their zero-velocity surfaces''
\citep{article--Ciotti_Dutta--1994--alignment}. As in
Sect.~\ref{sec:dynamical-app}, (i) we start with an unperturbed spherical
galaxy, (ii) we take an external tidal potential and add it to the (unperturbed)
galaxy potential, thus ignoring the changes in the galaxy potential induced by
the deformation of the galaxy \citep[see
also][]{article--Ciotti_Dutta--1994--alignment,
article--Bertin_Varri--2008--construction} , and (iii) we assume that the galaxy
immediateyl reacts to a change of the external gravity field by modifying its
shape accordingly (see Sect.~\ref{sec:dynamical-app} for details).

\subsection{General case}

In the absence of external fields, the galaxy potential $U_0$ obeys the Poisson
equation
\begin{equation}
\label{eq:poisson}
\nabla^2U_0=4\pi G\rho_0\;,
\end{equation}
where $\rho_0$ is the unperturbed galaxy mass distribution. We now
introduce a (weak) external potential $U_\mathrm{ext}$, so that
$U_0\mapsto U=U_0+U_\mathrm{ext}$.
The introduction of the external field changes the equipotential surfaces of the galaxy.
Given a volume $\mathcal V_0$ enclosed by a particular equipotential surface
$\partial\mathcal V_0=\{\forall\vec x|U_0(\vec x)=E_0\}$ at energy
$E_0$ of the unperturbed potential, we consider the corresponding equipotential surface
$\partial\mathcal V$ for $U$
\begin{equation}
        \label{eq:new-iso-surface}
        \partial\mathcal V=\{\forall\vec x|U(\vec x)=U_0(\vec x)+U_\mathrm{ext}(\vec
        x)=E_0+\delta E=E\}\;.
\end{equation}
The energy shift $\delta E$ is chosen in such a way that the
mass inside the surface does not change:
\begin{equation}
\label{eq:condizione-iso}
        \int\Theta\Big(E_0-U_0(\vec x)\Big)\rho_0(\vec x)\,\mathrm d^3\!x=\int\Theta\Big(E-U(\vec x)\Big)\rho_0(\vec
        x)\,\mathrm d^3\!x\;.
\end{equation}
A Taylor expansion of the right-hand side of
Eq.~\eqref{eq:condizione-iso} for low $\delta E-U_\mathrm{ext}$ gives
\begin{equation}
\int\delta\Big(E_0-U_0(\vec x)\Big)\rho_0(\vec x)\Big(\delta
E-U_\mathrm{ext}(\vec x)\Big)\,\mathrm d^3\!x=0\;.
\end{equation}
Since $\rho_0$ has spherical symmetry, $\rho_0$ and $\nabla U_0$ are uniform
on $\partial \mathcal V_0$ , and we finally obtain
\begin{equation}
\label{eq:delta-E}
\delta E=\frac{\oint_{\partial\mathcal V_0}U_\mathrm{ext}(\vec
x)\rho_0(\vec x)\bigl|\nabla U_0(\vec
x)\bigr|^{-1}\mathrm d^2\!x}{\oint_{\partial\mathcal V_0} \rho_0(\vec x)
\bigl|\nabla U_0(\vec x)\bigr|^{-1}\mathrm d^2\!x} 
= \frac{\oint_{\partial\mathcal V_0}U_\mathrm{ext}(\vec
x)\,\mathrm d^2\!x}{\oint_{\partial\mathcal
V_0}\mathrm d^2\!x}\;.
\end{equation}
If $U_\mathrm{ext}\equiv U_\mathrm{tidal}=\Phi_{ij}x_ix_j$, that is, the external potential is a
tidal one, $\delta E$ is equal to zero, because
$\Phi_{ij}$ is traceless and $\partial \mathcal V_0$ has spherical symmetry.

\begin{figure}
   \centering
   \begin{tikzpicture}[scale=1.5]
   \draw[->] (-1.8,0) to (1.8,0);
      \node at (1.6,0.2) {$\hat x_1$};
      \draw[->] (0,-1.5) to (0,1.5);
      \node at (0.2,1.5) {$\hat x_2$};
      \draw[dotted] (0,0) circle [radius=1];
      \draw (0,0) ellipse [x radius=1.2,y radius=0.8];
      \node at (0.2,1) {$\delta_2$};
      \node at (0.2,-1) {$\delta_2$};
      \node at (1.0,0.2) {$\delta_1$};
      \node at (-1.0,0.2) {$\delta_1$};
      \draw[very thick] (-1,0) -- (-1.2,0);
      \draw[very thick] (1,0) -- (1.2,0);
      \draw[very thick] (0,0.8) -- (0,1);
      \draw[very thick] (0,-0.8) -- (0,-1);
   \end{tikzpicture}
   \caption{Schematic representation of the deformation of a two-dimensional
   contour of an equipotential surface. The dotted line is the unperturbed
   contour, the solid line the perturbed contour. In the tidal approximation,
   the variations along the positive and negative directions are the same.}
   \label{fig:iso-curves}
\end{figure}
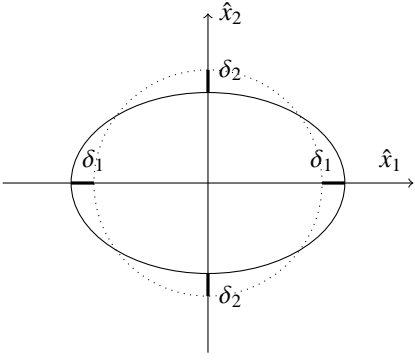

We use the galaxy logarithmic potential of Eq.~\eqref{eq:self-field}.
If we introduce an external tidal potential, the equipotential surface becomes
an ellipsoid (see Fig.~\ref{fig:iso-curves}). We place the galaxy center at the
origin and align the coordinate axes along the eigenvectors of the tidal tensor
$\Phi_{ij}$.  Then, we can evaluate the deviation from the circular shape of a
particular equipotential surface with radius $r_\mathrm{max}<r_\mathrm{tr}$ by
expanding to first order its semi-axis variations $\delta_i$ in
Eq.~\eqref{eq:new-iso-surface},
\begin{equation}
        \label{eq:general-dx}
        \delta_i=-\frac{\Phi_{ii}r_\mathrm{max}^2r_\mathrm{tr}^2+r_\mathrm{tr}^2\delta
        E}{MG(r_\mathrm{tr}+r_\mathrm{core})/(r_\mathrm{max}+r_\mathrm{core})+2r_\mathrm{max}r_\mathrm{tr}^2\Phi_{ii}}\;.
\end{equation}
We kept for generality $\delta E$, even if it vanishes for an external tidal
field. Equation~\eqref{eq:general-dx} allows us to compute the intrinsic ellipticity
of a particular isophotal of a galaxy subject to a tidal field as observed along any direction. For this,
we just have to project an ellipsoid with semi-axes $r_\mathrm{max}+\delta_1$,
$r_\mathrm{max}+\delta_2$ and $r_\mathrm{max}+\delta_3$ along the line of sight.
For example, if the ellipsoid is observed along $\vec{\hat x_3}$, we would have
from Eq.~\eqref{eq:def-ellipticity}
\begin{equation}
\label{eq:eps-general}
|\varepsilon^\mathrm s|=\frac{|\delta_1-\delta_2|}{2r_\mathrm{max}}\;.
\end{equation}

\subsection{Particular cases}
For the Keplerian tidal field \eqref{eq:tidal-kepler}, placing the external
spherical galaxy at $\vec R=(R,0,0)$, using \eqref{eq:eps-general}, and assuming $r_\mathrm{max},r_\mathrm{tr}\ll R$, we
obtain (Fig.~\ref{fig:eps})
\begin{equation}
\label{eq:epsilon-iso-kepler}
|\varepsilon^\mathrm{s}|\simeq\frac34\frac{M_\mathrm{ext}}{M}\frac{r_\mathrm{max}r_\mathrm{tr}^2}{R^3}\frac{r_\mathrm{max}+r_\mathrm{core}}{r_\mathrm{tr}+r_\mathrm{core}}\simeq1.6\times10^{-5}\;,
\end{equation}
where in the last step we used $M_\mathrm{ext}=\unit[10^{11}]{M_\sun}$,
$r_\mathrm{max}=\unit[10]{kpc}$ and $R=\unit[500]{kpc}$.
We note that $\varepsilon^\mathrm s=\mathop{\mathrm O}(R^{-3})$, and that the ellipticity
becomes lower for inner equipotential surfaces.

For an external DM filament directed along the line of sight and distant
$R=\unit[500]{kpc}$ from the galaxy, if we assume\footnote{With the values used in this paper, $2\lambda
  r_\mathrm{max}r_\mathrm{tr}^2 (r_\mathrm{core}+r_\mathrm{max})
  \simeq 10^{-2}MR^2 (r_\mathrm{core}+r_\mathrm{tr})$, and our
  assumption is justified.}
\begin{equation}
M\frac{r_\mathrm{tr}+r_\mathrm{core}}{r_\mathrm{max}+r_\mathrm{core}}\gg2\lambda \frac{r_\mathrm{max}r_\mathrm{tr}^2}{R^2}\;,
\end{equation}
we obtain from Eqs.~\eqref{eq:U-tidal-linear}, \eqref{eq:lambda} and
\eqref{eq:eps-general} (Fig.~\ref{fig:eps})
\begin{equation}
        \label{eq:eps-iso-linear}
   |\varepsilon^\mathrm s|\simeq\frac{\lambda r_\mathrm{max}r_\mathrm{tr}^2}{MR^2}
   \frac{r_\mathrm{max}+r_\mathrm{core}}{r_\mathrm{tr}+r_\mathrm{core}}
   \simeq4.9\times10^{-3}\;.
   \end{equation}
We observe that in this case the dependence on the distance is $\varepsilon^\mathrm
s=\mathop{\mathrm O}(R^{-2})$; again, inner
equipotential surfaces have lower ellipticity.

The values obtained through the equipotential approximation
\eqref{eq:epsilon-iso-kepler} and \eqref{eq:eps-iso-linear} are higher than
those obtained through the distribution function method \eqref{eq:eps-f-value}
and \eqref{eq:eps-f-linear}; the reason is that with the equipotential
approximation we only consider an outer isopotential surface, which is
more deformed than the inner ones.  Instead with the distribution function
method, more realistically, we are ``weighting the ellipticities of the
isophotes'' of the galaxy from its center to $r_\mathrm{max}$.
Equations~\eqref{eq:epsilon-iso-kepler} and \eqref{eq:eps-iso-linear} have the
same dependence on $R$ as was found with the distribution function method (see
Fig.~\ref{fig:eps}).

\end{document}